\documentclass[aps,prl,twocolumn,groupedaddress,showpacs,amsmath,amssymb]{revtex4}

\usepackage{graphicx}
\usepackage{dcolumn}
\usepackage{bm}

\begin{document}

\preprint{APS/123-QED}

\title{Spin echo in spinor dipolar Bose--Einstein condensates}

\author{Masashi Yasunaga}
\author{Makoto Tsubota}%
\affiliation{%
Department of Physics, Osaka City University, Sumiyoshi-ku, Osaka 558-8585, Japan
}%


\date{\today}

\begin{abstract}
We theoretically propose and numerically realize spin echo in a spinor Bose--Einstein condensate (BEC). We investigate the influence on the spin echo of phase separation of the condensate. The equation of motion of the spin density exhibits two relaxation times. We use two methods to separate the relaxation times and hence demonstrate a technique to reveal magnetic dipole--dipole interactions in spinor BECs. 
\end{abstract}

\pacs{03.75.Mn, 03.75.Nt }
\keywords{ spin-3 Bose--Einstein Condensate, Spin Echo, Magnetic Resonance, Magnetic Dipole--Dipole interaction}
\maketitle


NMR and ESR have revealed states of magnetic spin systems \cite{Slichter}. Especially the spectrum broadening contains informations of spin relaxation given by spin-orbit interaction, spin-spin interaction and inhomogeneous magnetic fields. Usually the inhomogeneity hides the other characteristic effects in the system. Spin echo technique, however, is able to remove effects of the  inhomogeneous diffusing. Thus the technique has been used in experiments of the magnetic spin system. 

The discovery of spin echo in 1950 by Hahn was momentous in the field of magnetic resonance  \cite{Hahn}. Hahn observed the recovery of a free induction decay signal of spins by using the combination of three $\pi/2$ pulses, naming the phenomenon ``spin echo''. Carr and Purcell later proposed other methods of producing spin echo. The first, called ``Method A'', uses the $\pi/2$--$\pi$ pulse sequence, while in ``Method B'' the echoes are obtained by applying many $\pi$ pulses after a $\pi/2$ pulse \cite{Carr}. Spin echo has contributed to the development of magnetic resonance, especially as a method of measuring the relaxation time for magnetic resonance imaging (MRI). MRI is used to visualize the structure of human bodies for medical examinations in hospitals. Spin echo is thus a field of physics with practical applications.

 In low temperature physics, particularly for superfluid $^3$He, NMR has made significant contributions in determining superfluid phases and revealing vortices and textures \cite{Vollhard}. For example, spin supercurrent \cite{Fomin, BR} and Bose--Einstein condensation of magnons\cite{Bunkov} have been still under investigation.  Spin echo also was found in A and B phases in the superfluid $^3$He \cite{Eska}.  Thus magnetic resonance in superfluid $^3$He has been developing.

Recently, BECs with a magnetic dipole--dipole interaction (MDDI) have received considerable attention \cite{LS, YK,HM,SY,MT}. The MDDI may be closely related to the relaxation mechanism of the spins. $^{52}$Cr BECs have been realized by Griesmaier {\it et al.} \cite{AG}. A $^{52}$Cr atom has a magnetic moment six times larger than that of an alkali atom. Hence, a $^{52}$Cr condensate clearly shows the effects of the MDDI as an anisotropic free expansion which depends on the orientation of the atomic dipole moments \cite{JS}. The magnitude of the MDDI can be controlled by modulating the s-wave scattering length \cite{TL}. Although the MDDI of alkali atomic BECs is quite small, it would be possible to observe the MDDI in the systems as well as in a $^{52}$Cr BEC \cite{YKK}. 
 
 
\begin{figure*}[t]
	\setlength\unitlength{\linewidth}
	\begin{picture}(1,0.3)
		\put(0.08,0.15){\includegraphics[trim= 15mm 15mm 15mm 15mm, clip, width=0.15\linewidth]{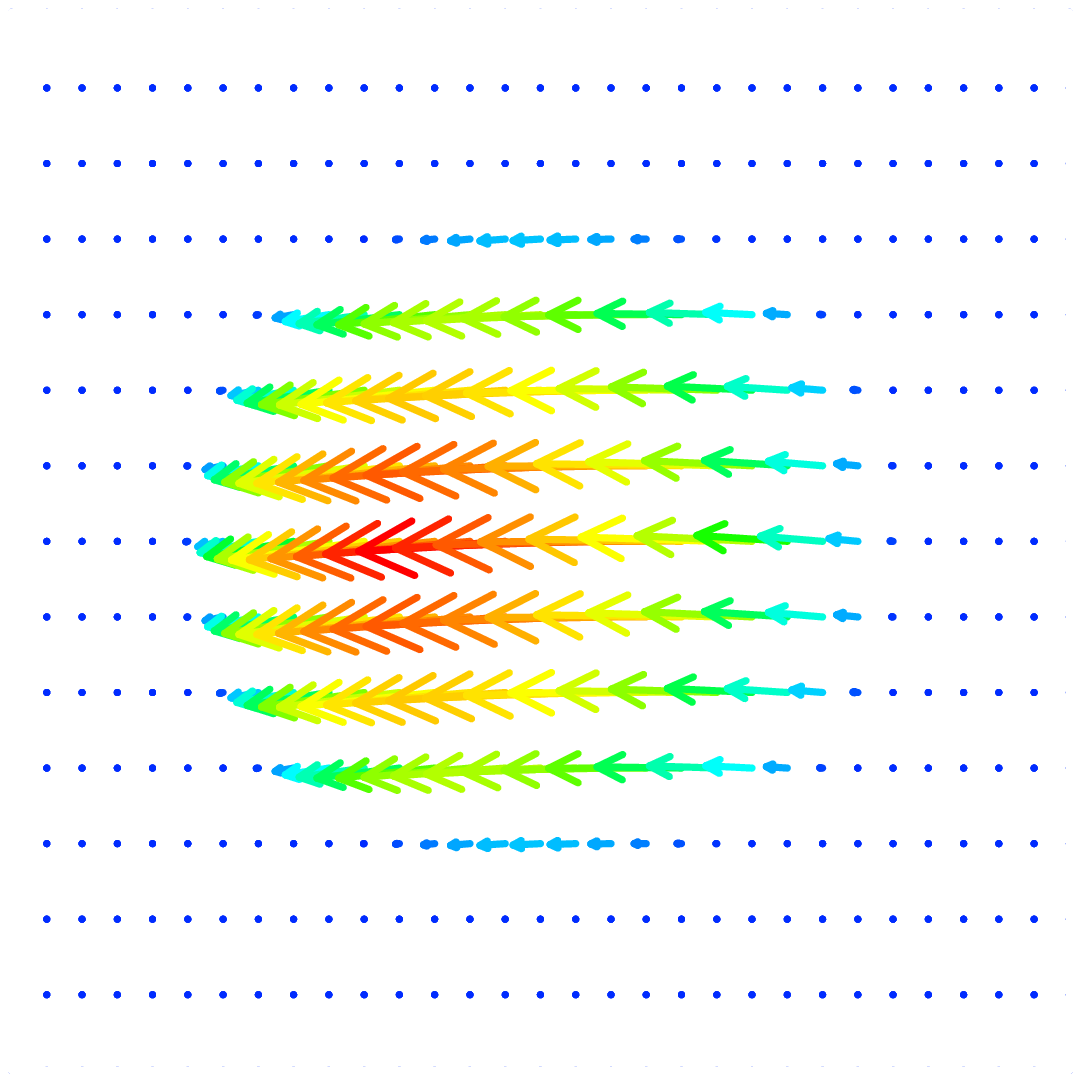}}
		\put(0.31,0.15){\includegraphics[trim= 15mm 15mm 15mm 15mm, clip, width=0.15\linewidth]{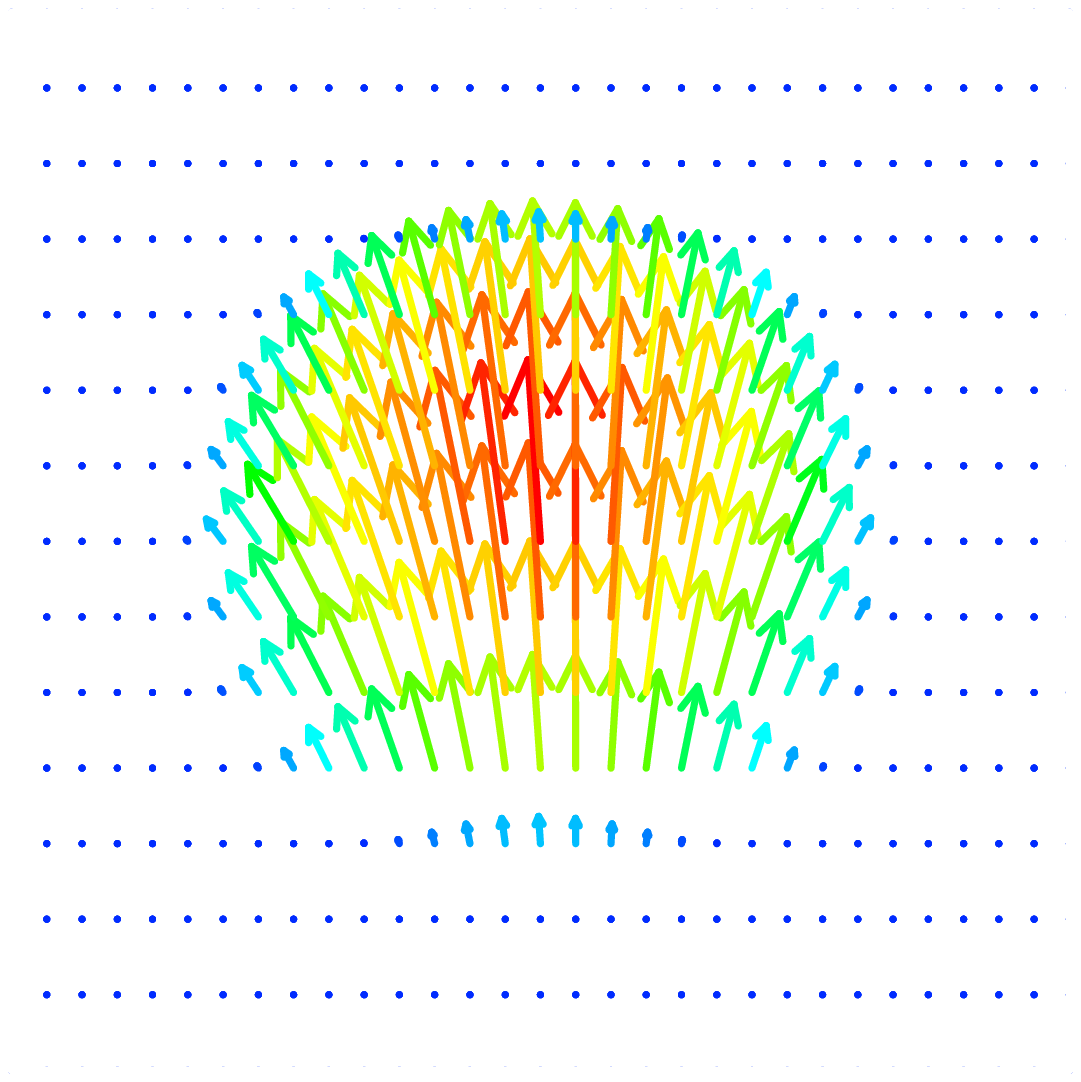}}
		\put(0.54,0.15){\includegraphics[trim= 15mm 15mm 15mm 15mm, clip, width=0.15\linewidth]{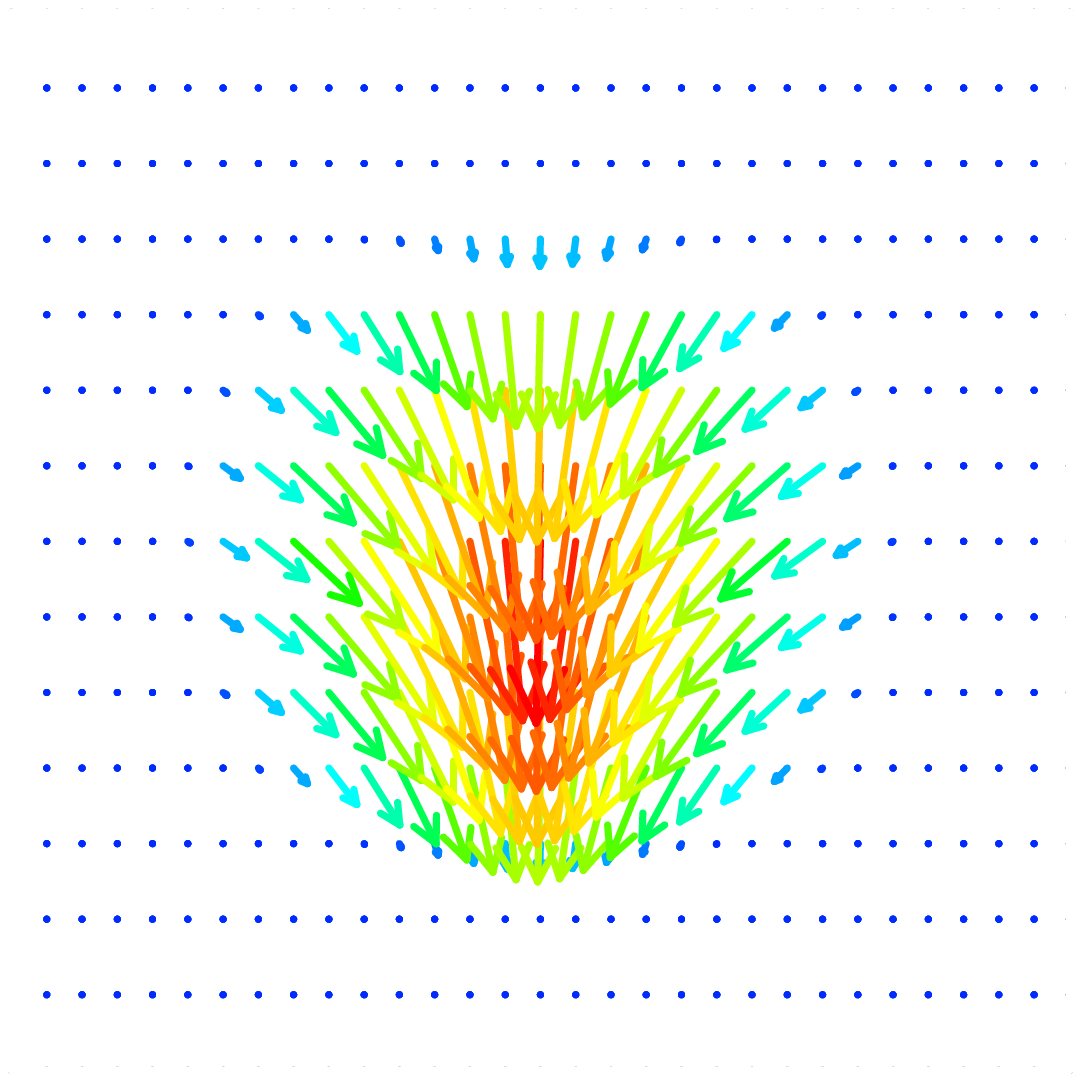}}
		\put(0.77,0.15){\includegraphics[trim= 15mm 15mm 15mm 15mm, clip, width=0.15\linewidth]{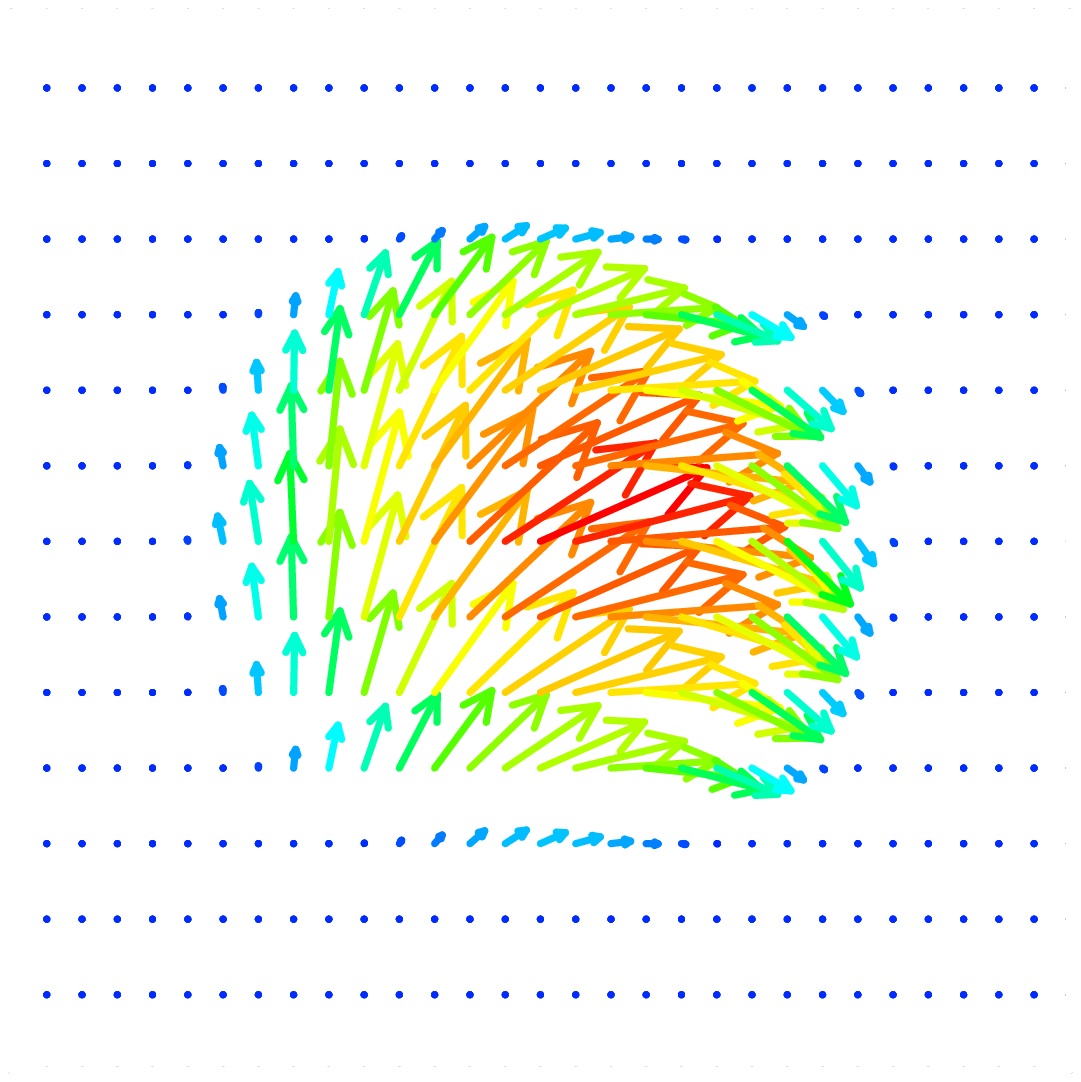}}
		\put(0.08,0){\includegraphics[trim= 15mm 15mm 15mm 15mm, clip, width=0.15\linewidth]{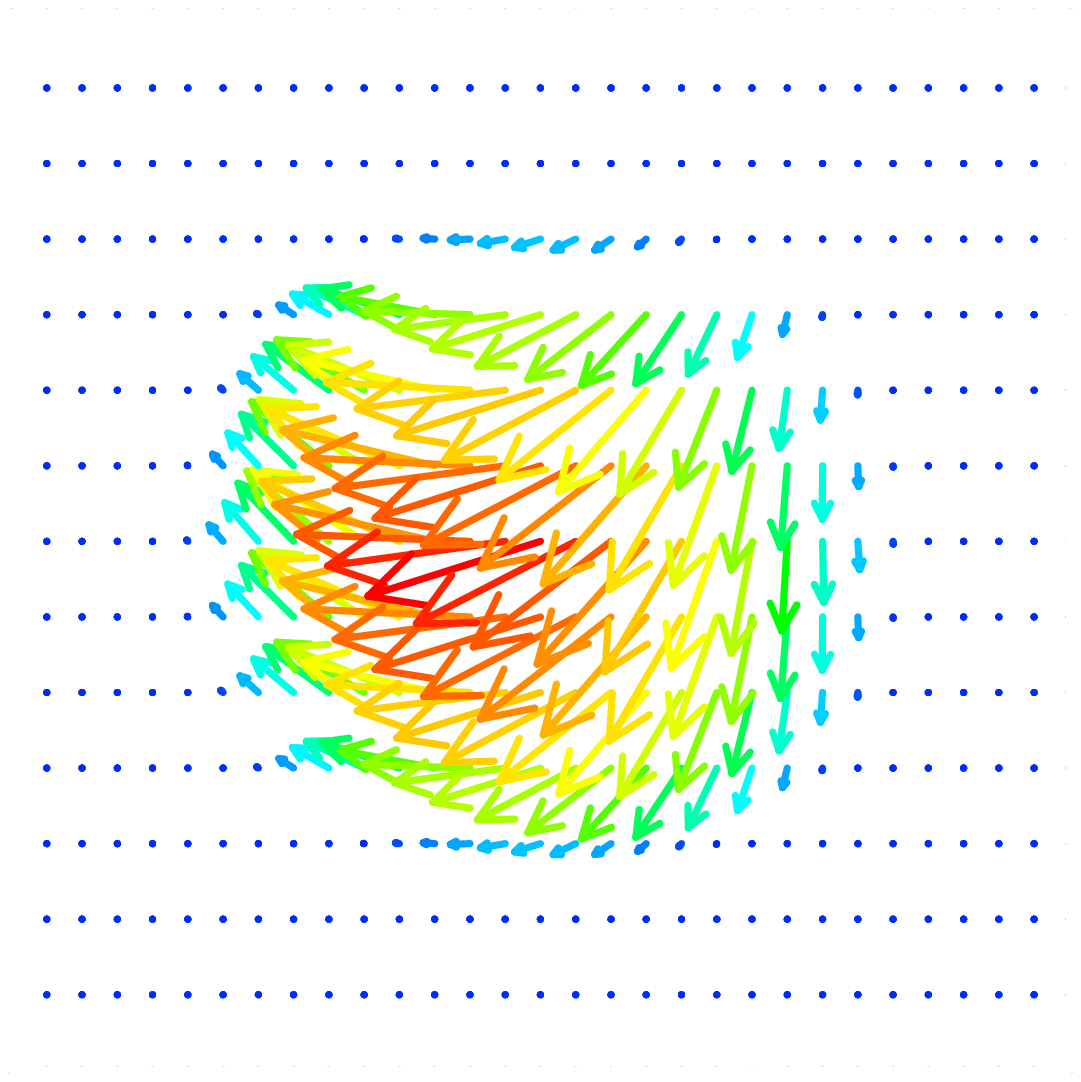}}
		\put(0.31,0){\includegraphics[trim= 15mm 15mm 15mm 15mm, clip, width=0.15\linewidth]{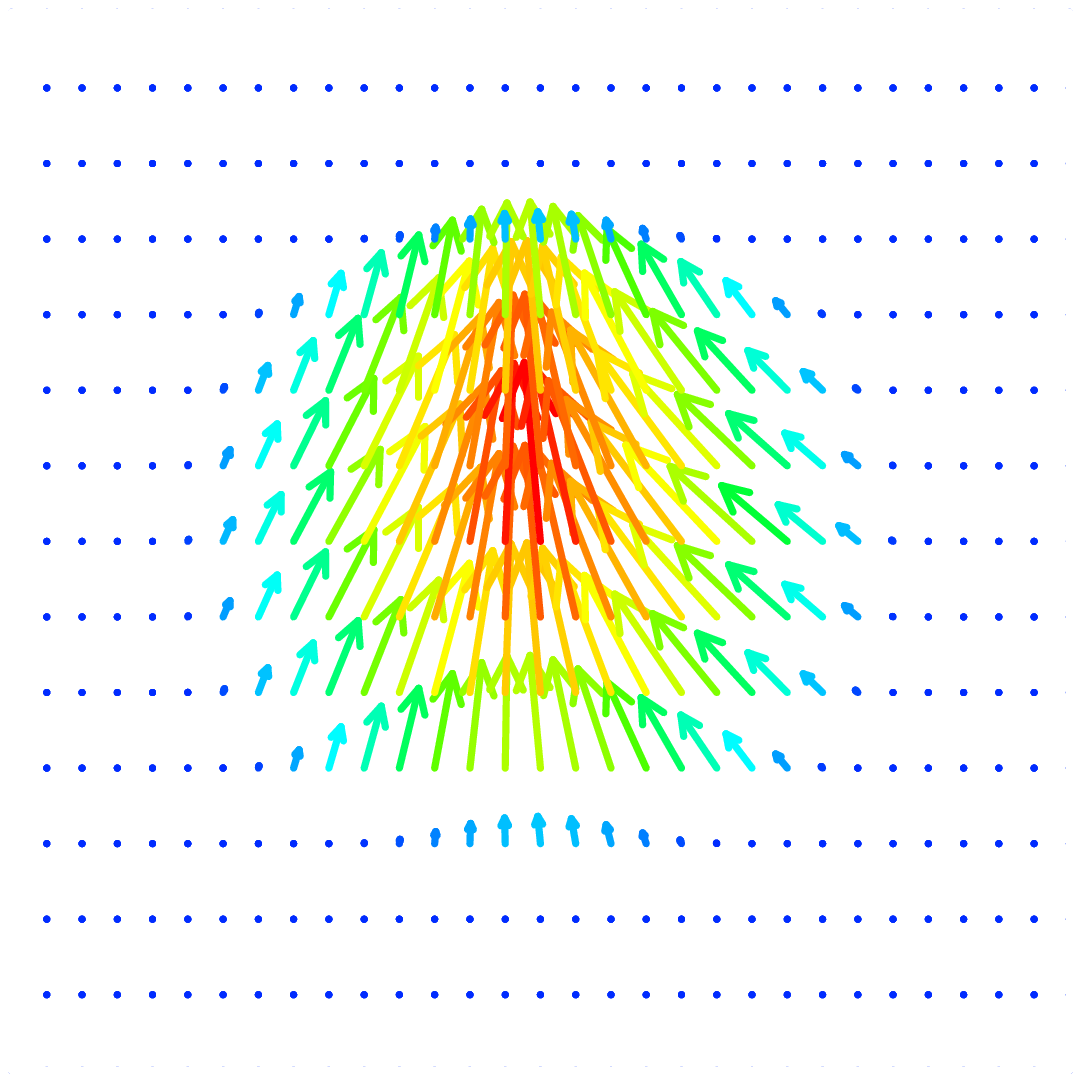}}
		\put(0.54,0){\includegraphics[trim= 15mm 15mm 15mm 15mm, clip, width=0.15\linewidth]{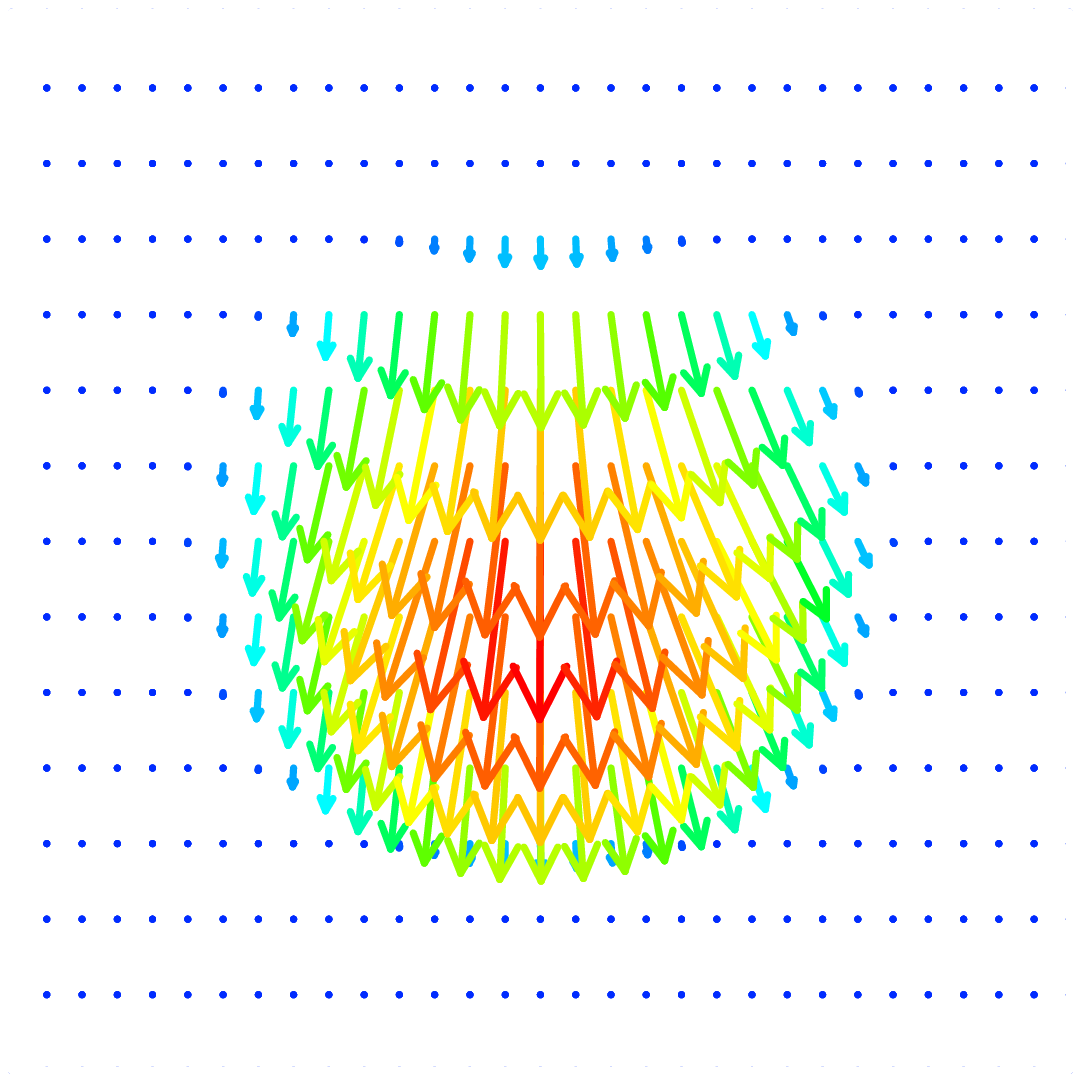}}
		\put(0.77,0){\includegraphics[trim= 15mm 15mm 15mm 15mm, clip, width=0.15\linewidth]{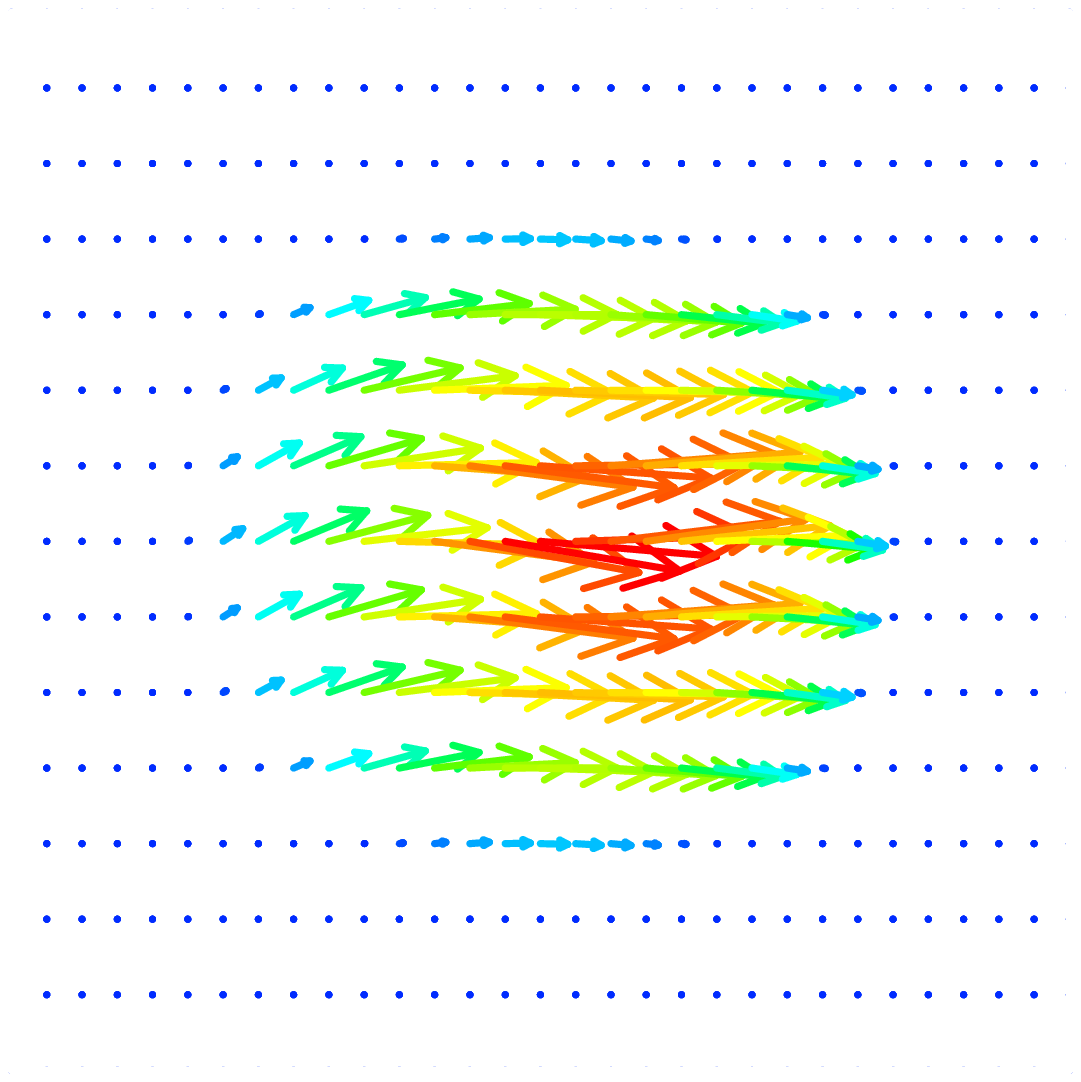}}		

		\put(0.09,0.15){\large{(a)} $\omega _{\perp}t = 0.75$}
		\put(0.32,0.15){\large{(b)} $\omega _{\perp}t = 2.4$}
		\put(0.55,0.15){\large{(c)} $\omega _{\perp}t = 4.1$}
		\put(0.78,0.15){\large{(d)} $\omega _{\perp}t = 5.8$}	
		\put(0.09,0.0){\large{(e)} $\omega _{\perp}t = 7.3$}
		\put(0.32,0.0){\large{(f)} $\omega _{\perp}t = 9.1$}
		\put(0.55,0.0){\large{(g)} $\omega _{\perp}t =10.6$}
		\put(0.78,0.0){\large{(h)} $\omega _{\perp}t = 12.8$}			
	\end{picture}
\caption{ (color online) Time development of the spin density vectors ${\bf S}$ projected on the $x$--$y$ plane in the spin echo with $\omega _{\perp}\tau= 5$. The magnitude of ${\bf S}$ increases from blue to red.}
\label{fig:dynamics}
\end{figure*}
 
In this letter, we propose spin echo in a spinor BEC for the first time, studying effect of Stern-Gelrach separation  and of MDDI by using Carr--Purcell--Meiboom--Gill sequence.  Investigating spin echo will provide information on the MDDI, which is exhibited even by $^{87}$Rb BECs.

First, we briefly review relaxation in magnetic resonance and spin echo. A spin {\bf S} with gyro-magnetic ratio $\gamma$ precesses with Larmor frequency $\omega _L = \gamma H_0$ under a homogeneous magnetic field ${\bf H} = H_0\hat{z}$, which is described by the equation of motion of the spin $d{\bf S}/dt = \gamma [{\bf S} \times {\bf H}]$. In many body systems of spin, if spins do not have an interaction with others, they precess with same Larmor frequency around a homogeneous magnetic field. On the other hand, The interaction between spins would make gradually the inhomogeneous precession through the local magnetic field. The decay signal, therefore, contains informations about spin-spin and spin-orbit interactions. However, if the magnetic field is inhomogeneous, the decay signal depends on also the inhomogeneity. Therefore it is difficult to choose effects of the interaction from the diffusion. Spin echo technique can remove the effect of  inhomogeneity. 

We consider that spins without interaction are polarized to the $z$ axis along an inhomogeneous magnetic field whose the spatial magnitude is almost distributed around $H_0$. Then by applying a $\pi/2$ pulse resonant with  $\gamma H_0$, the spins tilt to the $y$ axis with precessing. After the pulse, the spins precess on the $x-y$ plane, losing the coherence gradually. Subsequently applying a $\pi$ pulse  at $\tau$ after the first pulse reverses the direction of spins, recovering the coherence gradually. The coherence recovers perfectly at $\tau$ after the second pulse. This method is called ``spin echo''. On the other hand, the procedure can not give a perfect recovery when there is interaction between spins. The imperfect recovery comes from not the inhomogeneous magnetic field but the interaction.

To obtain spin echo in a $^{87}$Rb BEC trapped by a harmonic potential of frequency $\omega_{\perp}$, we study the two-dimensional spin-1 Gross--Pitaevskii (GP) equations with the MDDI:
\begin{eqnarray}\label{eq:GP}
i\hbar\frac{\partial \psi_\alpha}{\partial t} &=& \left(-\frac{\hbar^2}{2M}\nabla^2+V_{trap}-\mu\right)\psi_\alpha\nonumber \\
&&-g\mu_BH_iS^i_{\alpha \beta}\psi_\beta+c_0\psi^*_\beta\psi_\beta\psi_\alpha+c_2S_iS_{\alpha \beta}^i\psi_\beta \nonumber \\
&&+c_{dd}\int d{\bf r}' \frac{\delta_{ij}-3e^ie^j}{|{\bf r}-{\bf r}'|^3}S_i({\bf r}')S_{\alpha \beta}^j\psi_\beta.  
\end{eqnarray}
Here $V_{trap} = M\omega_{\perp}^2(x^2+y^2)/2$ is the trapping potential, $\mu$ the chemical potential, $g$ the Land$\grave{{\rm e}}$'s g-factor, $\mu_B$ the Bohr magneton, and 
\begin{equation}\label{eq:S}
S_i = \sum _{\alpha,\beta}\psi^*_\alpha S_{\alpha\beta}^i\psi_\beta
\end{equation}
 is the component of the spin density vector ${\bf S}$ represented by the spin matrices $S_{\alpha \beta}^i$. The short range interaction constants are $c_0 = 4\pi\hbar^2(a_0+2a_2)/3M$ and $c_2 = 4\pi\hbar^2(a_2-a_0)/3M$ given by the s-wave scattering lengths $a_0$ and $a_2$. The long range interaction constant is $c_{dd} = \mu_0g^2\mu_B^2/4\pi$ with vacuum magnetic permeability $\mu_0$.  For $^{87}$Rb, the parameters satisfy the relation $c_{dd} < -c_2\ll c_0$ \cite{SY}. We investigate the spin echo under a gradient magnetic field ${\bf H} = (Gx+H_0)\hat{z}$ with constant $G$.

 The dynamics start from  the stationary states with $|c_{dd}/c_2| = 0$ and $0.02$ in a gradient magnetic field of magnitude $0.98H_0 < |{\bf H}| < 1.02H_0$. Applying a ${\bf H}_{rot} = H_1\{\cos(\gamma H_0t)\hat{x}-\sin(\gamma H_0t)\hat{y}\}$ with $H_1 = 0.1H_0$, we calculate the GP equation using the Crank--Nicolson method, obtaining the spin echoes and effects of dephasing in the echo.

\begin{figure}[t]
\begin{center}
\includegraphics[width=0.9\linewidth]{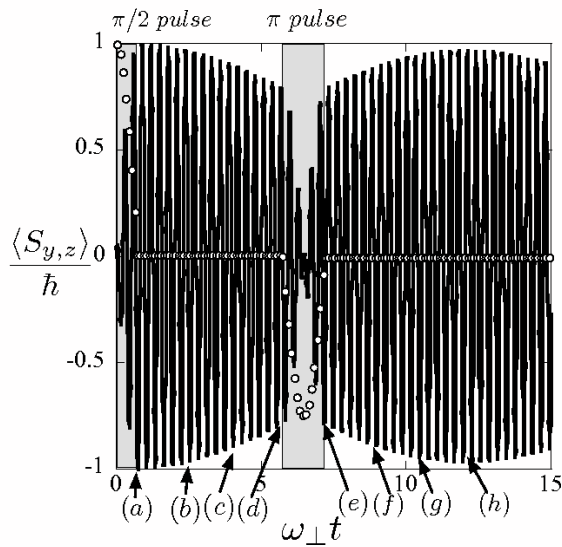} 
\caption{Time development of $\langle S_y\rangle/\hbar$ (black line) and $\langle S_z\rangle/\hbar$ (white circles). The symbols (a)--(h) indicate the times shown in Fig. \ref{fig:dynamics}. The gray zones represent the intervals of the pulses. The behavior of $\langle \hat{S}_z \rangle/\hbar$ shows that the $T_1$ relaxation is irrelevant to the dynamics $\langle \hat{S}_y \rangle/\hbar$.   }
\label{fig:signal}
\end{center}
\end{figure}

Spin echo is clearly observed in Fig. 1, showing the dynamics of ${\bf S}$ projected on a $x$--$y$ plane for $c_{dd} = 0 $ and $\omega _\perp \tau = 5$.  After the $\pi/2$ pulse, the spins are oriented in the -$y$ axis (Fig. \ref{fig:dynamics}(a)) and start precessing on the plane. Then, the precessions gradually diffuse because the Larmor frequency is spatially dependent due to the gradient magnetic field (Fig. \ref{fig:dynamics}(b), (c)). At $t = \tau+t_{\pi/2} $ (Fig. \ref{fig:dynamics}(d)), the $\pi$ pulse is applied, which reverses the direction of the spin (Fig. \ref{fig:dynamics}(e)). The spins then gradually become coherent (Fig. \ref{fig:dynamics}(f), (g)), eventually refocusing at $t_{peak} = 2\tau+t_{\pi/2}+t_{\pi}$(Fig. \ref{fig:dynamics}(h)). After the echo, the spins start to defocus again. These features are represented in the time development of the expectation value $\langle \hat{S}_{y, z} \rangle = \int d{\bf r} \psi_\alpha^*S_{\alpha\beta}^{y, z}\psi_\beta$ in Fig. \ref{fig:signal}. The signal of $\langle \hat{S}_y \rangle$ gradually decays from $ t_{\pi/2}$ (a) to $\tau+t_{\pi/2}$ (d) and then increases from $\tau+t_{\pi/2}+t_{\pi}$ (e) to $2\tau+3t_{\pi/2}$ (h). 

\begin{figure}[t]
\begin{center}
\includegraphics[width=0.9\linewidth]{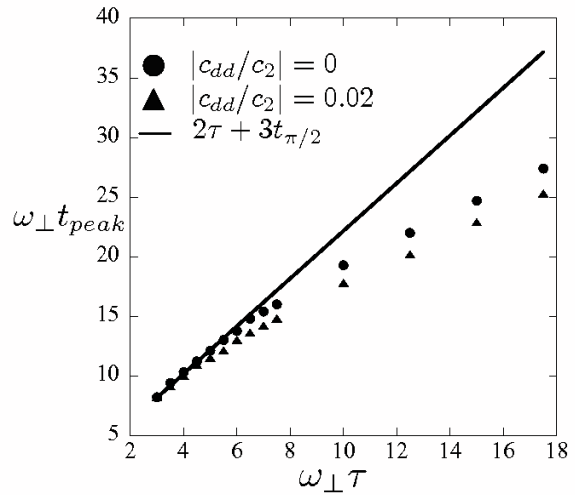} 
\caption{ Time $\omega _{\perp}t_{peak}$ at which echoes appear versus the interval $\omega _{\perp}\tau$ between $\pi/2$ and $\pi$ pulses. The circles and triangles are the results for $|c_{dd}/c_2| = 0$ and $0.02$, respectively.}
\label{fig:peaks}
\end{center}
\end{figure}

To show that this is genuine spin echo, we investigate $t_{peak}$ by changing $\tau$, thus confirming the spin echo for $|c_{dd}/c_2| = 0$ and $0.02$. Figure \ref{fig:peaks} indicates that the relation is satisfied until $\omega _\perp \tau = 6.5$ for $c_{dd} = 0$ and $4.5$ for $|c_{dd}/c_2| = 0.02$. These deviations from the relation $t_{peak} = 2\tau +3t_{\pi/2}$ can be understood by  investigating the equation of motion of spin based on the GP equations. From Eqs. (\ref{eq:GP}) and (\ref{eq:S}), we obtain the equation of motion
\begin{equation}\label{eq:torque}
\frac{\partial S_i}{\partial t} = \nabla\cdot{\bf J}_i+\gamma\left[ {\bf S} \times {\bf H}\right]_i+\frac{c_{dd}}{\hbar}R_{dd}^{j}S_{l\neq\{i,j\}},
\end{equation}
where ${\bf J}_i = \hbar/(2Mi)S_{\beta \alpha}^i(\psi_\beta^*\nabla\psi_\alpha-\psi_\alpha\nabla\psi_\beta^*)$ is term of spin current, which is similar to spin current in superfulid $^3$He \cite{Leggett},  and $R_{dd}^{j} = \int d{\bf r}' \frac{\delta_{jk}-3e^je^k}{|{\bf r}-{\bf r}'|^3}S_k({\bf r}')$. The first, second, and third terms of Eq. (\ref{eq:torque}) are derived from the kinetic, Zeeman, and MDDI terms of the GP equation. The differences between the circles and triangles in Fig. \ref{fig:peaks} are due to the dipole term, while their deviations from $t_{peak} = 2\tau +3t_{\pi/2}$ are caused by the spin current given by separation of the condensates due to the gradient magnetic field. 

\begin{figure}[t]
\begin{center}
\includegraphics[width=0.99\linewidth]{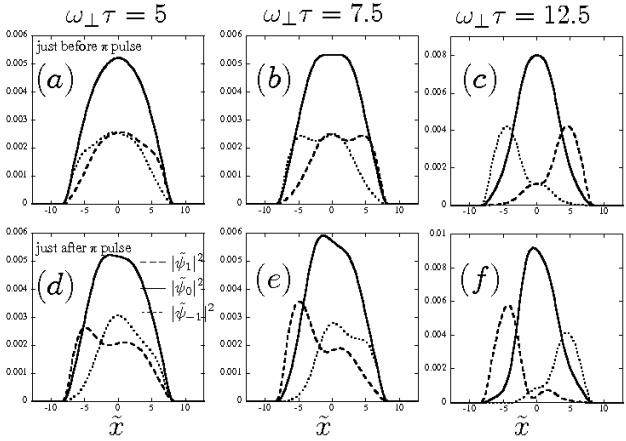} 
\caption{  Density profiles $|\tilde{\psi}_\alpha(y=0)|^2 = a_h^2|\psi_\alpha(y=0)|^2/N$ with  $a_h = \sqrt{\hbar/M\omega_\perp}$ and $\tilde{x} = x/a_h$ without the MDDI just before the $\pi$ pulse ((a), (b), (c)) and just after the $\pi$ pulse ((d), (e), (f)) for $\omega _\perp \tau = 5$, $7.5$, and $12.5$.}
\label{fig:dens}
\end{center}
\end{figure}

The magnetic resonance of this system is greatly affected by separation of spinor condensates. As shown in Fig. \ref{fig:dens} (a), (b), (c), the condensates separate under the gradient field after the $\pi/2$ pulse, keeping approximate symmetry between $x > 0$ and $x < 0$. Applying the $\pi$ pulse changes the $\psi_1$ and $\psi_{-1}$ components to $\psi_{-1}$ and $\psi_1$ respectively, which reverses {\bf S}. The $\pi$ pulse breaks the symmetry of the distribution of the condensates. After the pulse, $\psi_{\pm 1}$ move to the center, tending to overlap each other for $\omega _\perp \tau < 7.5$. The overlap is not clear for $\omega _\perp \tau > 7.5$. The effect is caused by a repulsive interaction between the different components, which is easily found in the GP equations of Eq. (\ref{eq:GP}). Thus, spin echo works properly for pulses shorter than $\omega_\perp \tau = 7.5$, because phase separation is not yet significant. On the other hand, the MDDI operates as an attraction between $\psi_1$ and $\psi_{-1}$ and makes the echo peaks appear earlier than the echo without the MDDI, as shown in Fig. \ref{fig:peaks}.

\begin{figure}[t]
\begin{center}
\includegraphics[width=0.99\linewidth]{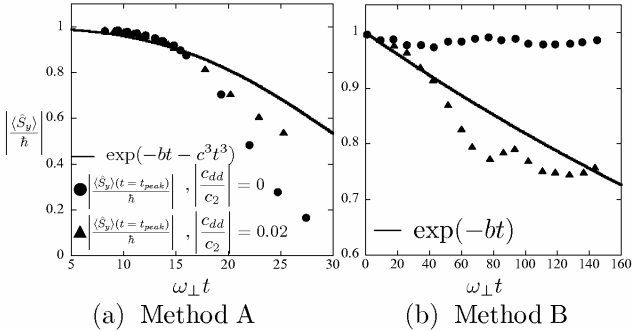} 
\caption{(a) Results of Method A, showing multiple exposures of $\langle \hat{S_y} \rangle/\hbar$ with (circles) and without (triangles) the MDDI at $ t_{peak}$ for various $\tau$, where $\omega _\perp\tau = 3.5, 4.0,\ldots, 7.5, 10, 12.5, 15,17.5 $. The black line shows the Carr--Purcell form $\exp[-bt-c^3t^3]$ with $b/c=0.072$. (b) Results of Method B, showing a single exposure of the signal of peaks with $\omega _\perp \tau = 3.5$ and $n = 17$. The line $\exp(-bt)$ is fit to the early peaks, which are little affected by $\exp(-c^3t^3/17^2)$. We obtain $\omega _\perp T_{dd} \sim 500$.  }
\label{fig:mAB}
\end{center}
\end{figure}

We now apply Methods A and B, which are major sequences in NMR and ESR for measuring two relaxation times, to spinor condensates. The dynamics of Eq. (\ref{eq:torque}) is characterized by two relaxation times. One is the relaxation time $T_{nu}$ due to the non-uniform condensates and the other is the dipolar relaxation time $T_{dd}$. In Method A, the decay of echo peaks is measured using a $\pi/2$--$\pi$ pulse sequence for various values of $\tau$. In liquids and gasses, the decay is generally caused by atomic self-diffusion; the first term of Eq. (\ref{eq:torque}) operates similarly in our systems. In Method B, a $\pi/2$--$\pi$--$\pi\ldots$ pulse sequence is used, with $n$ times $\pi$ pulses applied after a $\pi/2$ pulse. Carr {\it et al.} introduced the effect of diffusion on liquids and gasses into the decay of a signal by using a random walk model and obtained transversal decay of the $\pi/2$--$\pi$ sequence $\exp[-t/T_2-\gamma^2 a^2G^2t^3/24n^2\tau]$\cite{Carr}, where $a$ is a fixed distance of the random walk. Thus, the diffusion decreases with increasing $n$. The diffusion is given by the first term of Eq. (\ref{eq:torque}) in our case. For investigating $T_{nu}$ and $T_{dd}$, we performed a numerical simulation of Methods A and B, shown in Fig. \ref{fig:mAB}.  Figure \ref{fig:mAB} (a) shows multiple exposures of $\langle \hat{S}_y \rangle/\hbar $ at $t = t_{peak}$ for various $\tau$ for Method A. The echo peaks decay with $\tau$. When the condensates separate for large $\tau$, the decay does not satisfy the exponential behavior. Thus, we fit the exponential formula to the echo peaks until $\omega _\perp \tau = 7.5$, obtaining a diffusion time of $\omega _\perp t = 40$ when $\langle \hat{S}_y \rangle/\hbar$ decays to $e^{-1}$. This diffusion time should be equal to $T_{nu}$. Figure \ref{fig:mAB} (b) shows a single exposure of the echo peaks for Method B with $\omega_\perp \tau = 3.5$ and $n = 17$. In the simulation, rather than applying the original $\pi/2$--$\pi$--$\pi\ldots$ sequence, we apply a $\pi/2$--$\pi$--$(-\pi)$--$\pi$--$(-\pi)\ldots$ sequence, considered by Meiboom and Gill \cite{SM}, to cancel deviation due to repeated $\pi$ pulses. In the case of no MDDI, the echo peaks do not decay, showing that the Method B sequence removes the effect of phase separation. While the signal with the MDDI decays, the relaxation time should be given by $\omega_\perp T_{dd} \sim 500$. These results show $T_{nu} < T_{dd}$. Method B enables us to reveal the MDDI through $T_{dd}$ in spinor dipolar BECs even if the MDDI is too weak to be easily detected.

In conclusion, we have numerically realized spin echo in a trapped $^{87}$Rb BEC by calculating the spin-1 two-dimensional GP equation. We have investigated how the spin echoes are affected by the phase separation and the dipole--dipole interaction. The equation of motion of the spin density (Eq. (\ref{eq:torque})) derived from the GP equation shows two relaxation times $T_{nu}$ and $T_{dd}$. In order to distinguish $T_{dd}$ from $T_{nu}$ by using spin echo, we introduce Methods A and B, obtaining the relation  $T_{nu} > T_{dd}$. The study of spin echo of spinor BECs will help develop the field of magnetic resonance in cold atomic BECs system and may be applicable to MRI in these systems. Although this work is limited to the two-dimensional case, three-dimensional spinor BECs, especially with the MDDI, should show interesting features depending on geometry. A study of three-dimensional spinor BECs will be reported soon elsewhere. We also would like to study spin wave and spin transport as analogy to magnetic resonance of superfluid $^3$He.    


M. Y. acknowledges the support of a research fellowship of the Japan Society for the Promotion of Science for Young Scientists (Grant No. 209928). M. T. acknowledges the support of a Grant-in Aid for Scientific Research from JSPS (Grant No. 18340109) and a Grant-in-Aid for Scientific Research on Priority Areas from MEXT (Grant No. 17071008).


\end{document}